\documentstyle[emulateapj]{article}

\begin{document}

\title{Correlation length of X-ray brightest Abell clusters}

\author{
Mario G. Abadi, \altaffilmark{2}
 Diego G. Lambas  \altaffilmark{1}
and Hern\'an Muriel \altaffilmark{1}}
\affil{Grupo de Investigaciones en Astronom\'{\i}a Te\'orica y Experimental
(IATE)}
\affil{Observatorio Astron\'omico de C\'ordoba, Laprida 854, 5000, C\'ordoba,
Argentina\\
mario@oac.uncor.edu,
dgl@oac.uncor.edu,
hernan@oac.uncor.edu} 

\altaffiltext{1}{CONICET, Argentina}
\altaffiltext{2}{present address, Physics Department, University of Durham, U.K.}

\begin{abstract}

We compute the cluster auto-correlation function $\xi_{cc}(r)$  
of an X-ray flux limited sample of Abell clusters (XBACs, \cite{ebe}). 
For the total XBACs sample we find a power-law fit 
$\xi_{cc}=(r/r_0)^{\gamma}$ with $r_0=21.1$ Mpc h$^{-1}$and $\gamma =-1.9$
consistent with the results of $R \ge 1 $ Abell clusters.
We also analyze $\xi_{cc}(r)$  for 
subsamples defined by different X-ray luminosity thresholds where we
find a weak tendency of larger values of $r_0$ with increasing
X-ray luminosity although with a low statistical significance.
In the different subsamples analyzed we find 
$21 < r_0 < 35 $ Mpc h$^{-1}$ and $-1.9< \gamma < -1.6$.
Our analysis suggests that
cluster X-ray luminosities may be used for a reliable  
confrontation of cluster spatial distribution
properties in models and observations.
 
\end{abstract}


\section{INTRODUCTION}

Different authors have analyzed the cluster-cluster spatial two-point
correlation function finding power-law fits of the form 
$\xi_{cc}(r)=(r/r_{0})^{\gamma}$  with $\gamma \simeq -1.8$ (\cite{bs83}, 
\cite{pw92}). 
The value of the cluster-cluster correlation 
length $r_0$ is  controversial, as also is its 
dependence on cluster mass.  
This has been achieved by studying samples
selected by cluster richness and the associated
mean inter-cluster separation  
$d_c=n_c^{-1/3}$ where $n_c$ is the mean number density of clusters.
\cite{bw92} and \cite{bc92}
argue for a universal scaling relation for the two-point correlation function of
rich clusters
where the cluster correlation length satisfies $r_{0}=0.4 d_c$.
At low values of $d_c \simeq 30-50 h^{-1}$ Mpc the APM Cluster Survey (\cite{dalton94}),
the Edinburgh/Durham Cluster Catalog \cite{lumsden92} (hereafter EDCC), and
the \cite{a58} and \cite {aco89} cluster samples give similar results
consistent with $r_0=15-20 h^{-1}$ Mpc. 
At larger $d_c$, however, the analyses rely
only on the Abell catalog and on a cluster sample selected from the APM
Galaxy Survey (\cite{croft97}). The results of this high richness APM cluster 
sample are not consistent with the universal scaling relation
derived from Abell clusters by \cite{bw92} since  
only a weak dependence of $r_0$ on $d_c$ is found in the rich APM cluster sample.
A partial explanation for the different results between rich Abell and APM 
clusters could rely on the fact that
the Abell catalog is subject to visual artificial inhomogeneities in
contrast with the automated and well controlled APM cluster catalog 
(\cite{croft97}). 
It should be noted, however, that given the steeper correlations for
richer clusters these results are not inconsistent
with the universal relation in terms of correlation amplitude. 

The problems of projection effects in cluster selection (see
\cite{van97}) 
may be strongly overcome by selecting clusters in the X-ray rather than the
optical. Moreover, given the good correlation between 
X-ray luminosity and cluster mass ($L_X \propto M^{{4 \over 3}}$)
found in both analytical solutions 
(\cite{bert85}) and in numerical simulations 
(\cite{nav95}),
an X-ray selected sample is suitable to study the dependence of cluster
spatial correlations on mass.  
In this work we explore the values of $r_0$  
 in subsamples taken from the X-ray brightest Abell-type clusters of galaxies, 
hereafter XBACs (\cite{ebe}).
This sample of clusters is complete in X-ray flux, and we have selected
subsets with different cuts in X-ray luminosity $L_x$.

\section{DATA AND ANALYSIS}

The X-ray-brightest Abell-type clusters of galaxies survey 
(hereafter XBACs, \cite{ebe}) comprises 277 objects and is 
 a 95 $\%$ complete flux limited sample.  
We have restricted this catalog to galactic latitudes $|b|>25^o $ 
and X-ray flux
 $f > f_{cut} = 5\times 10^{-12}$ erg cm$^{-2}$ s$^{-1}$ in the $0.1-2.4$ keV band 
comprising a final sample of 248 clusters.
This sample although optically selected is confirmed by the
X-ray emission of the intra-cluster gas thus excluding spurious Abell clusters
generated by projection effects. Also, as discussed by \cite{ebe}, the XBACs
sample is unaffected to first order by the incompleteness in volume of the 
Abell catalog at large distances since missing Abell clusters
of low richness would not be included in XBACs due to their low X-ray luminosity.

In figure 1 we plot the X-ray luminosity of the clusters $L_x$ as a 
function of redshift $z$ taken from Table 3 of \cite{ebe}.
The smooth curve corresponds to the luminosity of an object with flux $f_{cut}$
at redshift $z$ in a flat cosmology.
Cluster distances $d$ were derived
using  the standard relation (e.g., \cite{sandage})  

\begin{equation}
d=\frac{c(q_0z+(q_0-1)((1+2q_0z)^{1/2}-1))}{h_0q_0(1+z)^2}
\end{equation} 

where $z$ is the cluster redshift, 
$h_0$ is the Hubble constant in units of 100 $km s^{-1} Mpc^{-1}$, 
and $c$ is the speed of light. 
Throughout this paper we have adopted a deceleration parameter $q_0=0.5$.

% fig 1 ************************************************
\placefigure{fig-1}
\input epsf
\def\figureps[#1,#2]#3.{\bgroup\vbox{\epsfxsize=#2
    \hbox to \hsize{\hfil\epsfbox{#1}\hfil}}\vskip12pt
    \small\noindent Figure#3. \def\par{\endgraf\egroup\vskip12pt}}
%\begin{figure*}[b]
\figureps[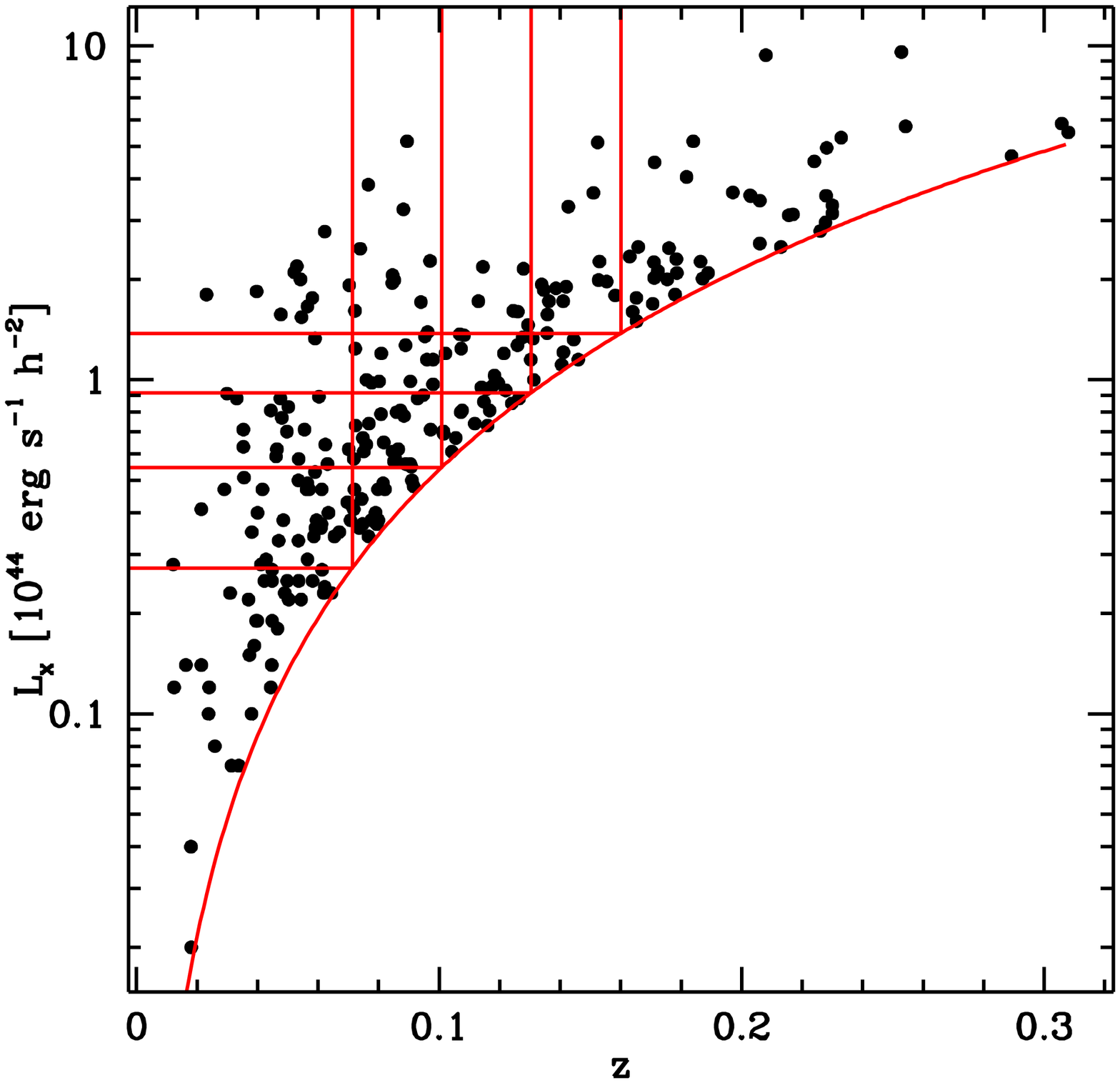,1.00\hsize] 1. X-ray luminosity $L_x$ vs redshift $z$ for the total
sample of of 248 clusters analyzed. The smooth curve displays the luminosity corresponding 
to flux $f_{cut}$ at redshift $z$ in a flat cosmology.
%\end{figure*}                                                          
%************************************************

We calculate the cluster-cluster two-point spatial correlation function
$\xi_{cc}(r)$
cross correlating the data with a random
catalog 
constructed by randomizing the angular positions of
the clusters with the same redshift distribution.
Each random catalog has $n_{ran}$ points homogeneously 
distributed within the same boundary 
than the subsamples.
To compute $\xi_{cc}(r)$ we have used the estimator 
\begin{equation}	
        \xi_{cc}(r)=2 f \frac{n(r)}{n_{ran}(r)}-1
\end{equation}
where $n(r)$ and $n_{ran}(r)$ 
are the 
number of cluster-cluster and cluster-random pairs separated 
by a distance $r$ respectively; $f=N_{ran}/(N-1)$ where $N$ and $N_{ran}$
are the total number of clusters in the observed sample and random catalog 
respectively.

We have considered 4 lower limits in X-ray luminosity
which  define volume incomplete subsamples of clusters
(subsamples 1i to 4i).
We have also defined other 4 subsamples by further imposing the restriction
of a cut in redshift ($z_{cut}$) in order to build volume complete subsamples
(subsamples 1c to 4c) (see tables 1 and  2). In figure 1 the horizontal
lines define the lower limits of the 4 incomplete subsamples.
The regions above the horizontal lines and with limiting redshift
at the vertical lines
define the 4 complete subsamples. The resulting number N of clusters
is between 85 and 214 in the incomplete subsamples, and between 43 and 72 in the
complete subsamples.                                          

% tab 1 **********************************************
\begin{tabular}{@{}ccccc@{}}
%\begin{deluxetable}{cccccc}[h]
%\tablewidth{35pc}
%\tablewidth{20pc}
%\tablecaption{Incomplete Subsamples}
%\tablehead{
%\colhead {Subsample} &
\hline
&&&& \nl
Sample & $L_x$ & N & $\gamma$ & $r_0$ \nl 
          & $10^{44}h^{-2} erg s^{-1}$ & & & Mpc $h^{-1}$  \nl 
%\colhead{ $L_x [10^{44} h ^{-2} erg \ s^{-1}$]  }&
%\colhead {N} &
%\colhead{ $\gamma$ }&
%\colhead{ $r_0$ [Mpc $h^{-1}$ ] }
% }
%\startdata
&&&& \nl
\hline
All& $\geq$ 0.02 & 248 & -1.92  & 21.1 $^{+ 1.6}_{- 2.3}$  \nl
1i & $\geq$ 0.27 & 214 & -1.89  & 22.1 $^{+ 2.7}_{- 2.9}$  \nl
2i & $\geq$ 0.54 & 168 & -1.80  & 23.5 $^{+ 4.5}_{- 4.7}$  \nl
3i & $\geq$ 0.91 & 117 & -1.59  & 30.1 $^{+ 8.8}_{-10.9}$  \nl
4i & $\geq$ 1.38 &  85 & -1.75  & 27.0 $^{+11.2}_{-16.0}$  \nl
%\enddata
%\end{deluxetable}                                 
\hline
\end{tabular}
\vskip .2cm
{\footnotesize{\centerline{Table 1. Incomplete Subsamples}}}
\vskip .5cm
%**********************************************

We have fitted the correlation functions obtained with power laws
of the form $\xi_{cc}(r)=(r/r_0)^{\gamma}$. We have estimated the
best fitting parameters $\gamma$ and $r_0$ and their associated 
errors using a 
a maximum-likelihood estimator using a $
\chi ^2-$minimization procedure developed by Levemberg and Marquard, (see
\cite{pre87}). 
This method deals with the errors in each distance bin
providing a reliable set of fitting parameters to the correlation function.
In our calculations we assume Poisson errors 
$\simeq \sqrt {n(r)}$ in each bin to estimate the
uncertainty in the correlation function (see \cite{croft97} and references therein
).

% tab 2 **********************************************
\vskip .5cm
\begin{tabular}{cccccc}
%\begin{deluxetable}{cccccc}[h]
%\tablewidth{35pc}
%\tablecaption{Complete Subsamples}
%\tablehead{
%\colhead {Subsample} &
%\colhead{ $L_x [10^{44} h ^{-2} erg \ s^{-1}$]  }&
%\colhead{ $z$  }&
%\colhead {N} &
%\colhead{ $\gamma$ }&
%\colhead{ $r_0$ [Mpc $h^{-1}$]  }
% }
\hline
 &&&&& \nl
Sample & $L_x [10^{44}$ & $z$ & N & $\gamma$ & $r_0$ \nl
          & $ h ^{-2} erg \ s^{-1}$] & ($\leq$) & & & Mpc $h^{-1}$ \nl

&&&& \nl
%\startdata
\hline
1c &$\geq$0.27 & 0.071 & 59 & -1.76  & 26.4 $^{+ 6.9}_{-7.9}$  \nl
2c &$\geq$0.54 & 0.101 & 72 & -1.80  & 24.6 $^{+ 6.5}_{-8.3}$  \nl
3c &$\geq$0.91 & 0.130 & 54 & -1.59  & 30.1 $^{+ 8.8}_{-10.9}$ \nl
4c &$\geq$1.38 & 0.160 & 43 & -1.77  & 34.7 $^{+12.3}_{-16.9}$ \nl
%\enddata
%\end{deluxetable}                             
\hline
\end{tabular}
\vskip .2cm
{\footnotesize{\centerline{Table 2. Complete Subsamples}}}
\vskip .5cm
%***********************************************

% fig 2 ******************************************************
\placefigure{fig-2}
\input epsf
\def\figureps[#1,#2]#3.{\bgroup\vbox{\epsfxsize=#2
    \hbox to \hsize{\hfil\epsfbox{#1}\hfil}}\vskip12pt
    \small\noindent Figure#3. \def\par{\endgraf\egroup\vskip12pt}}
%\begin{figure*}[h]
\figureps[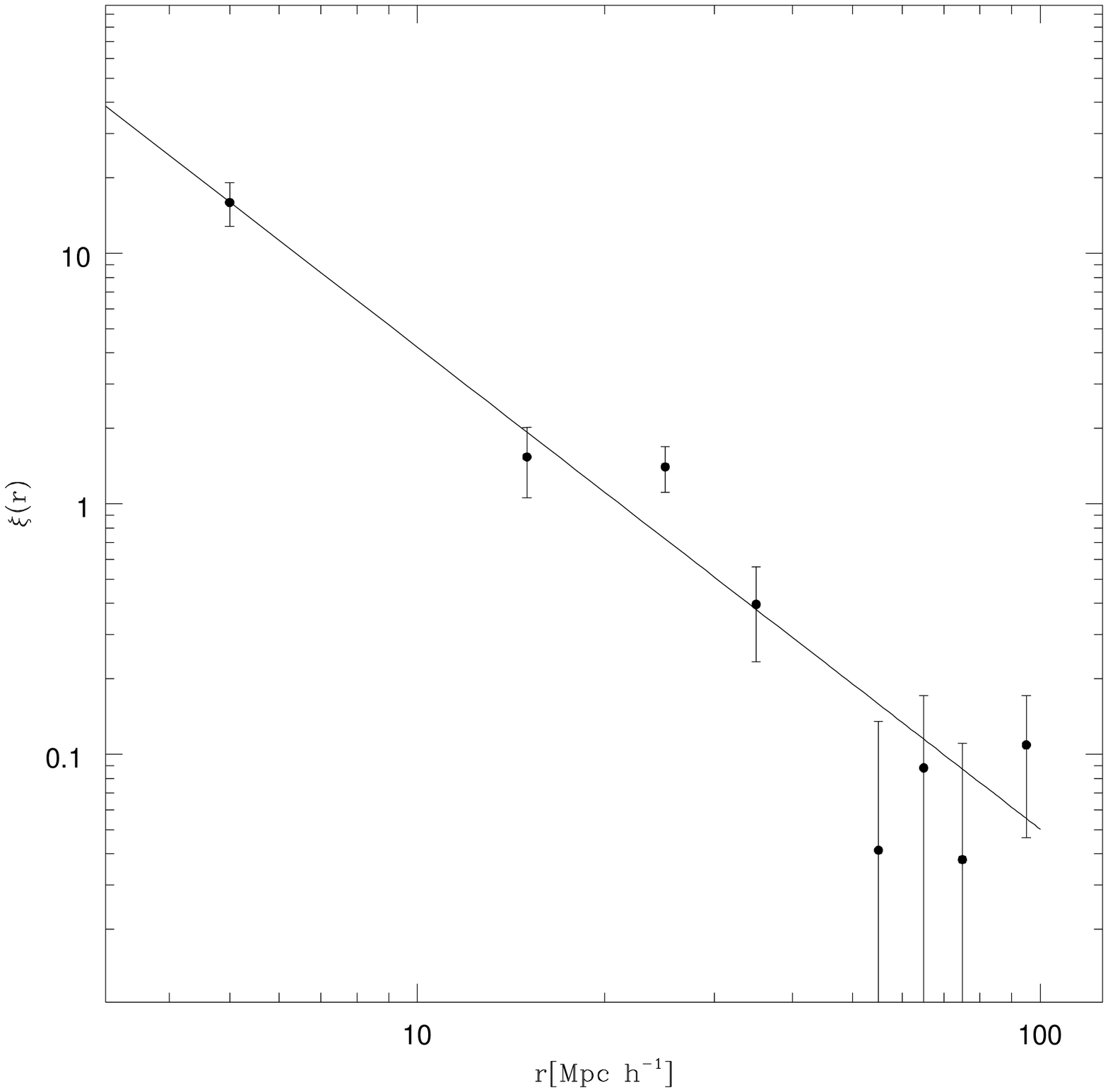,1.00\hsize] 2. Cluster-cluster two-point correlation functions 
$\xi_{cc}(r)$ corresponding to  the total sample.
%\end{figure*}                                                          
%*************************************************************

In figure 2, 3a and 3b are shown $\xi_{cc}(r)$ corresponding to the total sample, 
the incomplete subsamples 1i-4i, and the complete subsamples 1c-4c respectively. 
Error bars in $\xi_{cc}(r)$ correspond to Poisson estimates of the uncertainties
in the number statistics $\simeq \sqrt {n(r)}$.

% fig 3a *****************************************************
\vskip -2.cm
\placefigure{fig-3a}
\input epsf
\def\figureps[#1,#2]#3.{\bgroup\vbox{\epsfxsize=#2
    \hbox to \hsize{\hfil\epsfbox{#1}\hfil}}\vskip12pt
    \small\noindent Figure#3. \def\par{\endgraf\egroup\vskip12pt}}
%\begin{figure*}[b]
\figureps[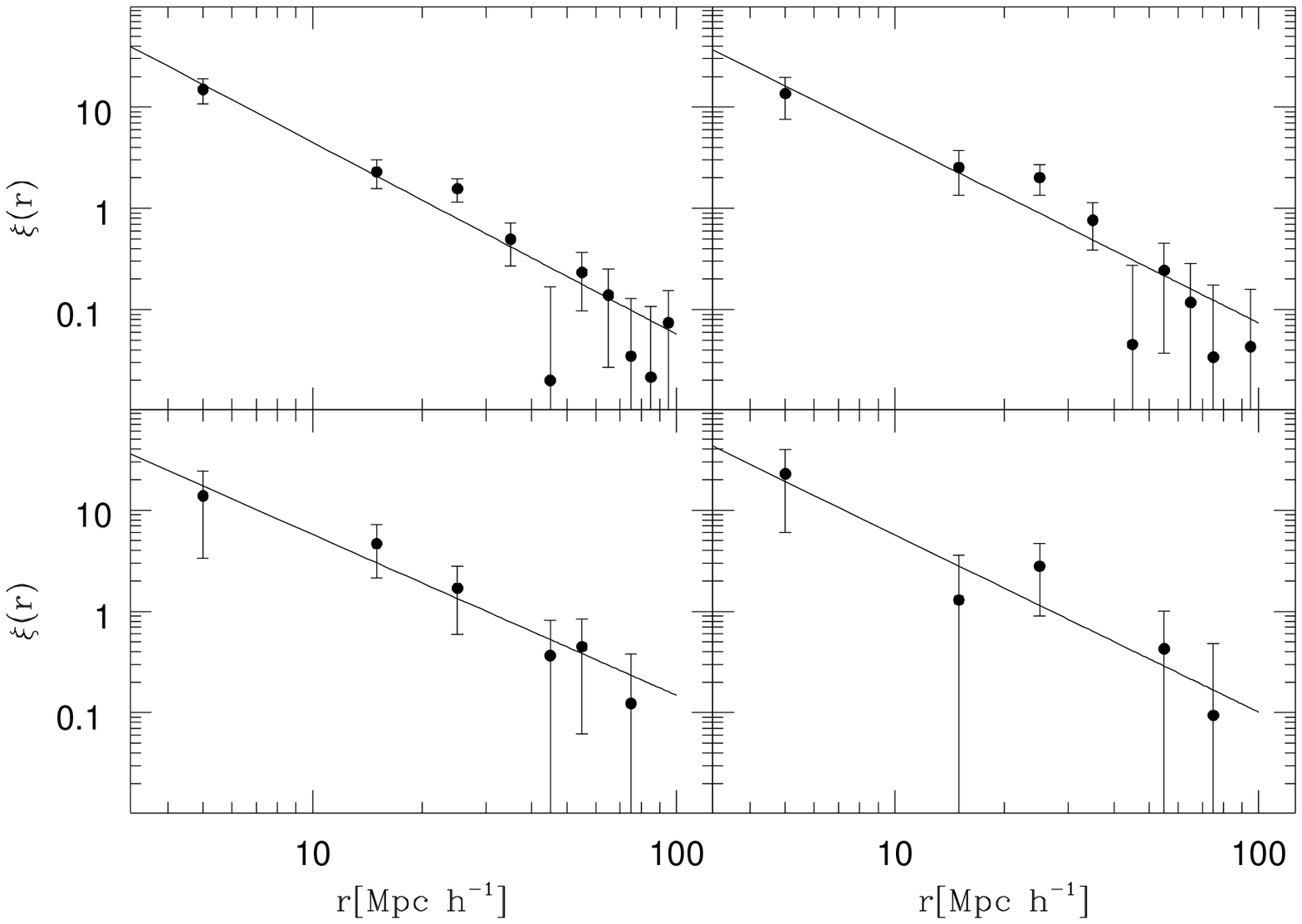,1.00\hsize] 3a. Cluster-cluster two-point correlation 
functions $\xi_{cc}(r)$ corresponding to  the different incomplete subsamples analyzed.  
%\end{figure*}                                                          
%*************************************************************

% fig 3b *****************************************************
\vskip -2.cm
\placefigure{fig-3b}
\input epsf
\def\figureps[#1,#2]#3.{\bgroup\vbox{\epsfxsize=#2
    \hbox to \hsize{\hfil\epsfbox{#1}\hfil}}\vskip12pt
    \small\noindent Figure#3. \def\par{\endgraf\egroup\vskip12pt}}
%\begin{figure*}[b]
\figureps[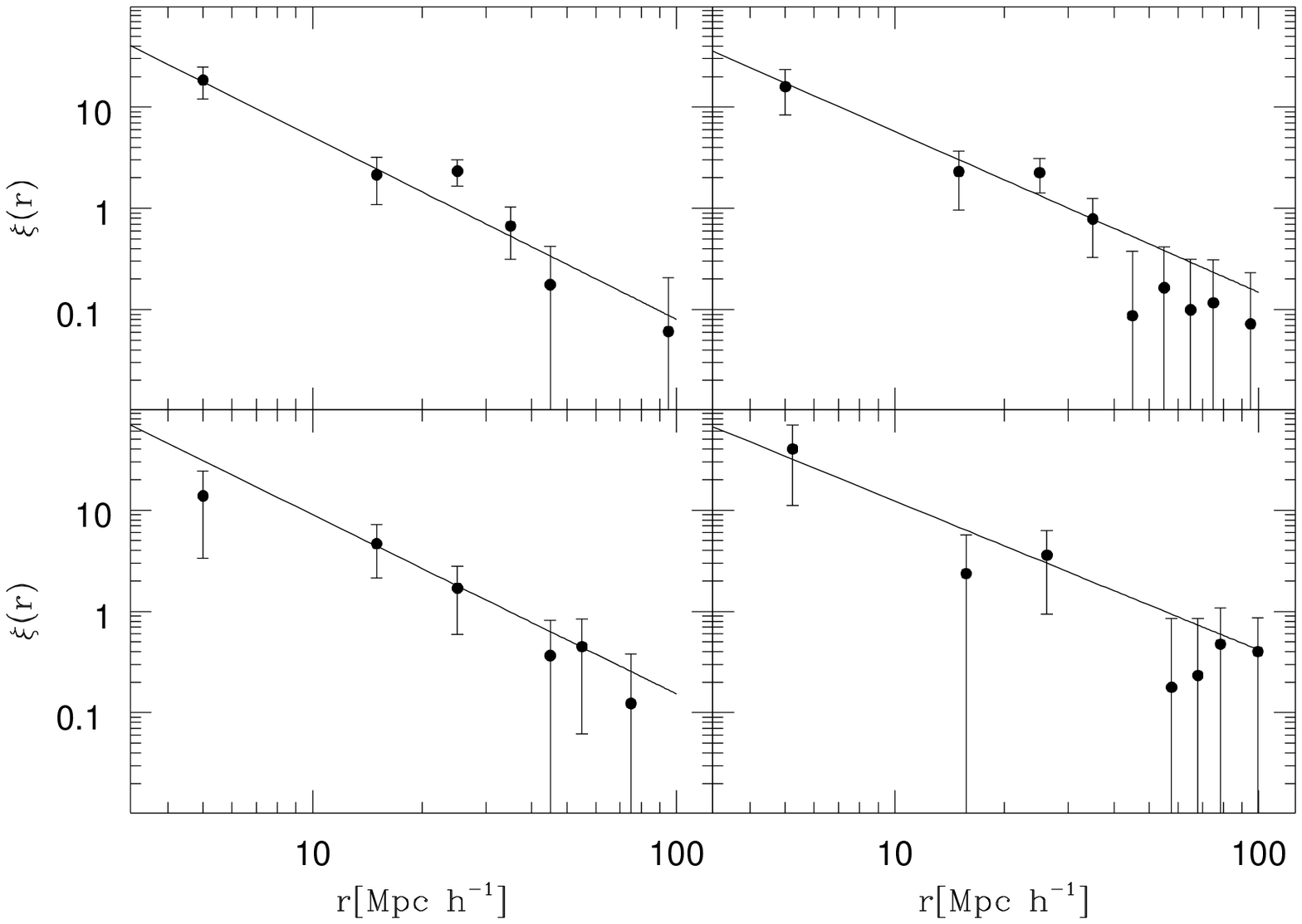,1.00\hsize] 3b. Cluster-cluster two-point correlation 
functions $\xi_{cc}(r)$ corresponding to  the different complete subsamples analyzed. 
%\end{figure*}                                                          
%*************************************************************

Estimates of the uncertainties in the power-law best-fitting parameters $r_0$ and 
$\gamma$
of the correlation functions may be visualized as plots of error contours 
$\chi^2 -\chi^2_{ML}$ in the $r_0 - \gamma$ plane.
In figures 4, 5a and 5b we show the corresponding error contours of 
confidence (1, 2 and 3 $\sigma$
level)  of the total sample,  
incomplete subsamples 1i-4i, and complete subsamples 1c-4c.

% fig 4 *********************************************************
\placefigure{fig-4}
\input epsf
\def\figureps[#1,#2]#3.{\bgroup\vbox{\epsfxsize=#2
    \hbox to \hsize{\hfil\epsfbox{#1}\hfil}}\vskip12pt
    \small\noindent Figure#3. \def\par{\endgraf\egroup\vskip12pt}}
%\begin{figure*}[b]
\figureps[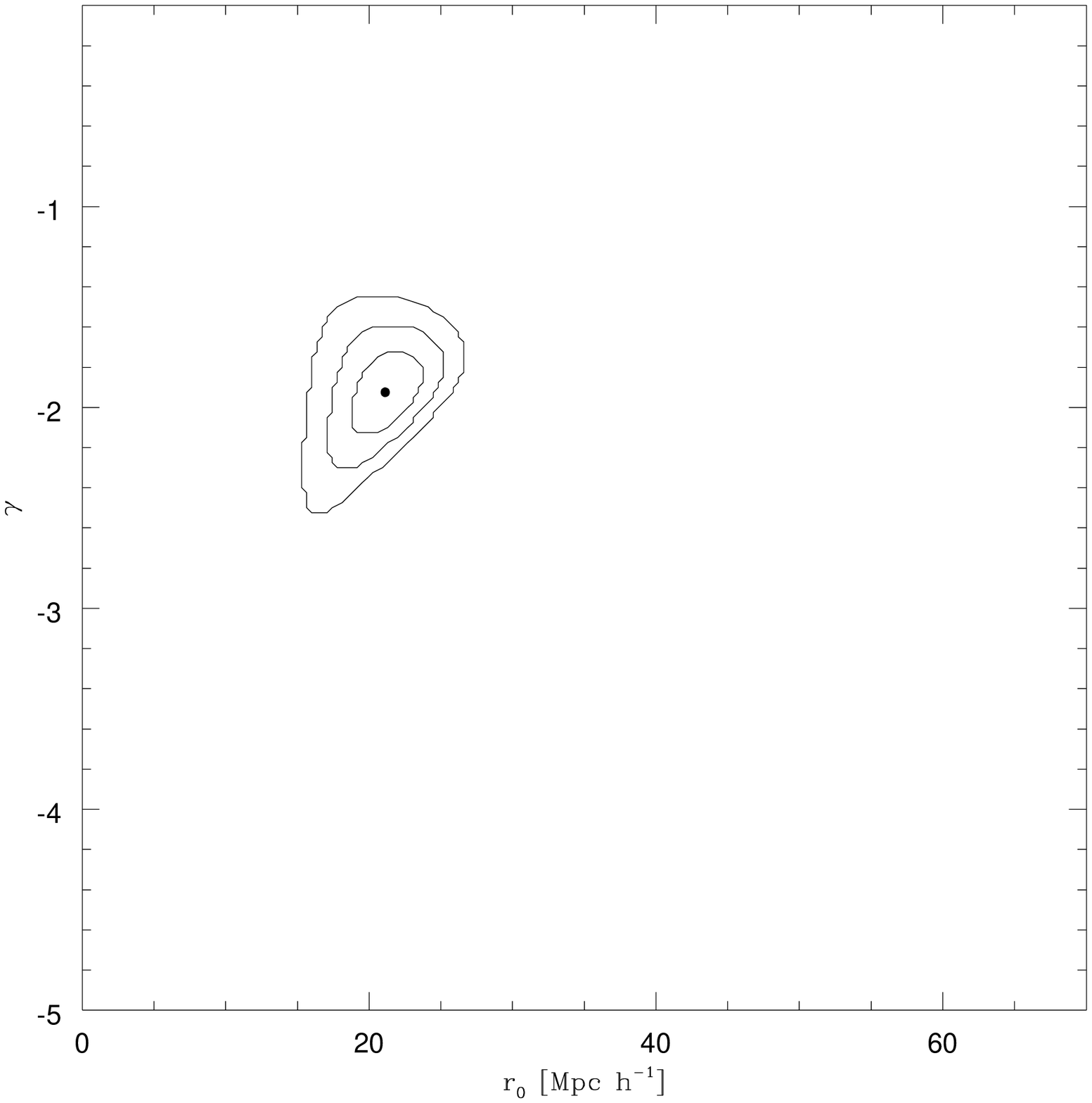,1.00\hsize] 4. Ellipses of confidence (1, 2 and 3 $\sigma$
level) corresponding to the correlation functions shown in figures 2.
%\end{figure*}                                                          
%**************************************************************

% fig 5a *********************************************************
\vskip -2.cm
\placefigure{fig-5a}
\input epsf
\def\figureps[#1,#2]#3.{\bgroup\vbox{\epsfxsize=#2
    \hbox to \hsize{\hfil\epsfbox{#1}\hfil}}\vskip12pt
    \small\noindent Figure#3. \def\par{\endgraf\egroup\vskip12pt}}
%\begin{figure*}[b]
\figureps[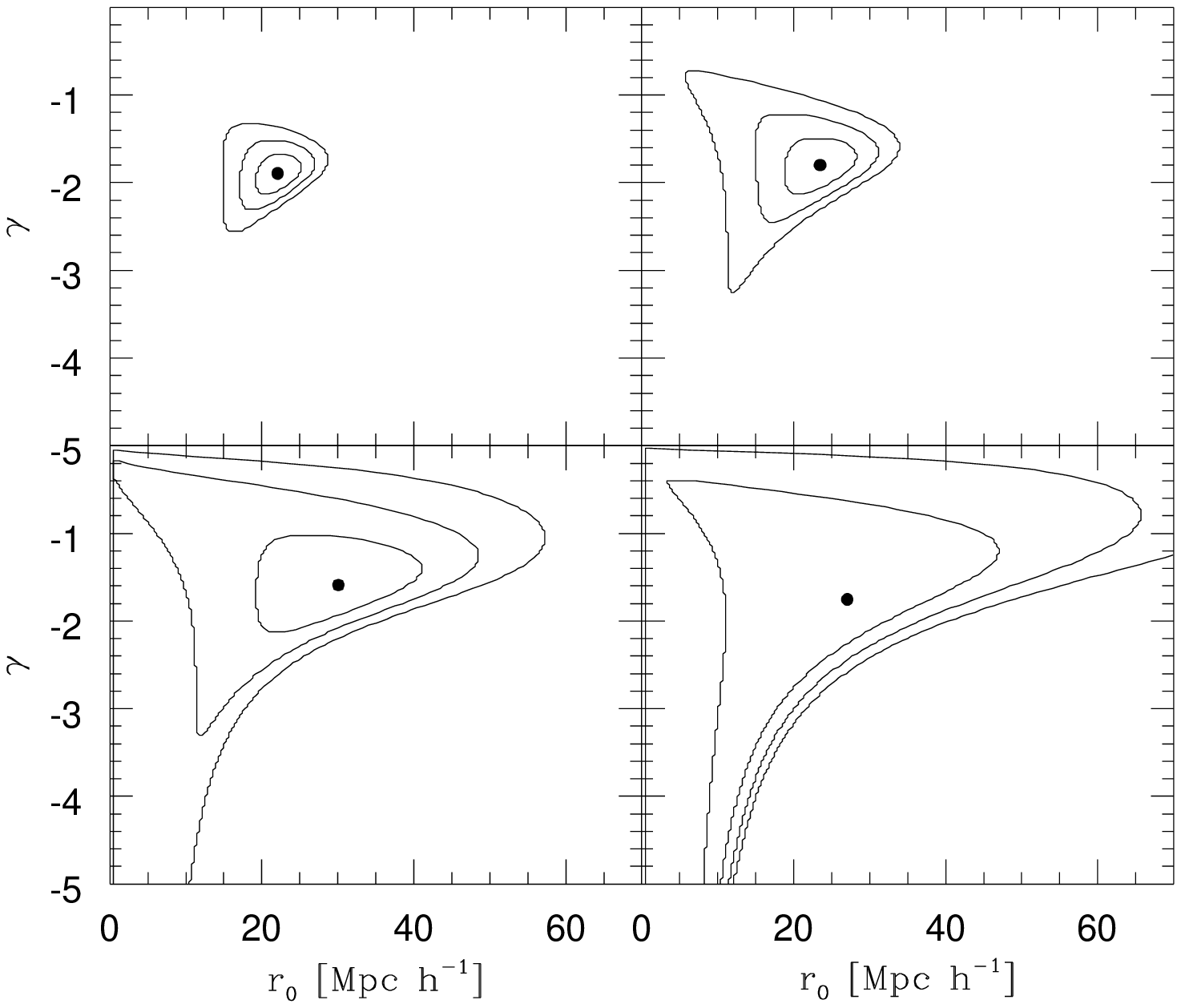,1.10\hsize] 5a. Ellipses of confidence (1, 2 and 3 $\sigma$
level) corresponding to the correlation functions shown in figures 3a.
%\end{figure*}                                                          
%**************************************************************

% fig 5b *********************************************************
\vskip -2.cm
\placefigure{fig-5b}
\input epsf
\def\figureps[#1,#2]#3.{\bgroup\vbox{\epsfxsize=#2
    \hbox to \hsize{\hfil\epsfbox{#1}\hfil}}\vskip12pt
    \small\noindent Figure#3. \def\par{\endgraf\egroup\vskip12pt}}
%\begin{figure*}[b]
\figureps[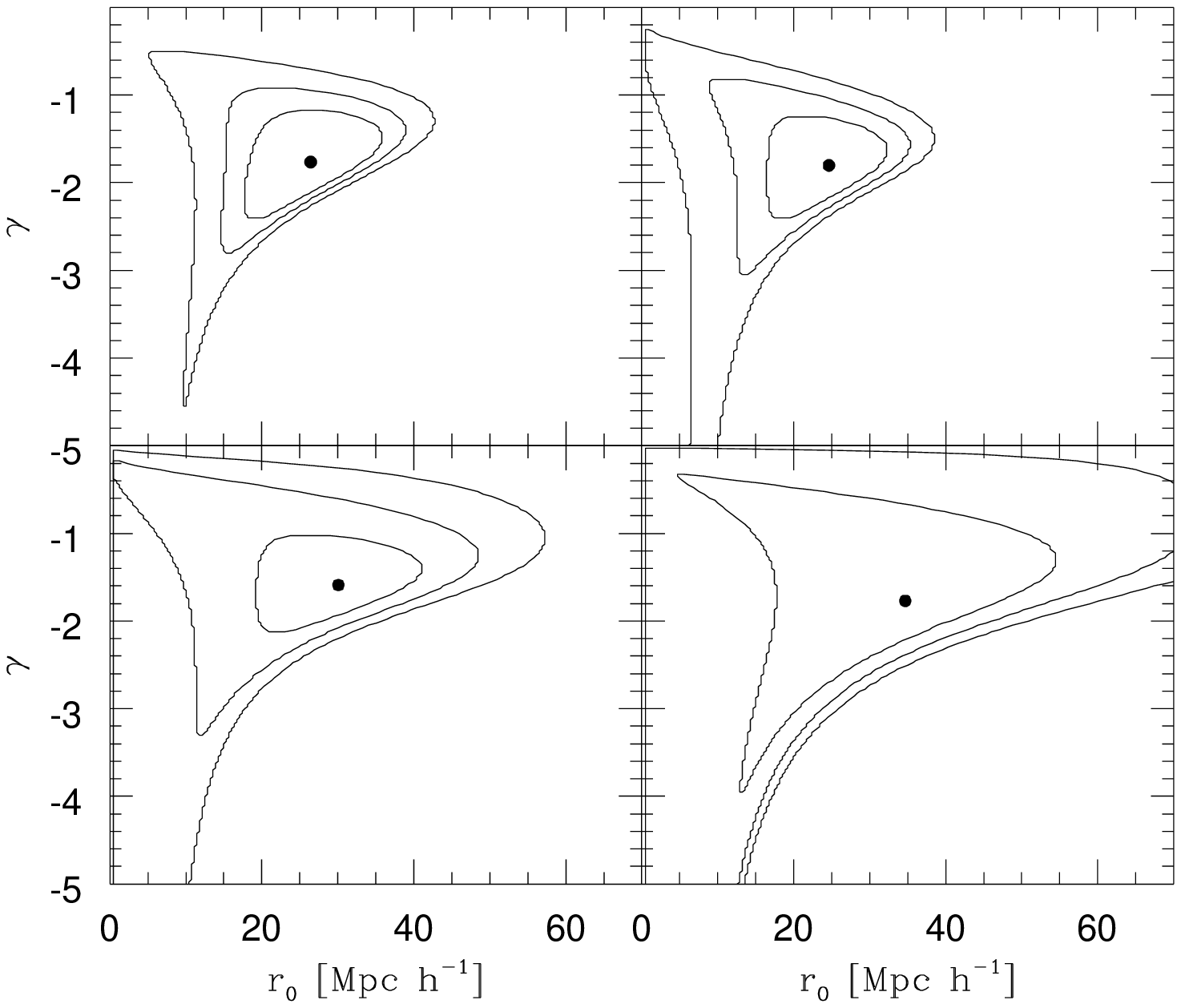,1.10\hsize] 5b. Ellipses of confidence (1, 2 and 3 $\sigma$
level) corresponding to the correlation functions shown in figures 3b.
%\end{figure*}                                                          
%**************************************************************

\section {CONCLUSIONS}

We have analyzed the two-point spatial correlation function
of clusters of galaxies selected from 
a sample of X-ray brightest 
Abell-type clusters of \cite{ebe}. 
For the total XBACs sample we find a power-law fit of the form 
$\xi_{cc}(r)=(r/r_0)^{\gamma}$ with $r_0=21.1 ^{+1.6}_{-2.3}$ Mpc $h^{-1}$ and 
$\gamma=-1.92$, values consistent with those derived for 
$R \ge 1$ Abell clusters (see \cite{bw92} and references therein).

In order to provide an insight of 
 the dependence of the cluster spatial correlation length $r_0$ on 
mass, we have estimated auto-correlation functions
for subsamples of clusters with different X-ray luminosity thresholds.
We find a weak increase of the correlation amplitude 
with increasing X-ray luminosity which is not statistically
significant and suggests a lack of a strong dependence of $r_0$ 
on cluster mass. For instance, in our complete subsample 4i with the highest
X-ray luminosity threshold $L_x > 1.38$  $10^{44} h^{-2} erg s^{-1}$
we obtain the highest value of correlation length, 
$r_0 \simeq 34.7^{+12.3}_{-16.9} $ Mpc h$^{-1}$.
Nevertheless, this value does not differ significantly from 
$r_0 \simeq 26.4^{+6.9}_{-7.9} $ Mpc h$^{-1}$
corresponding to subsample 1c with
$L_x > 0.27 10^{44} h^{-2} erg s^{-1}$.
 
There is a well documented evidence for the dependence of the correlation 
length on cluster richness as indicated
by the relation between $r_0$ and the mean inter-cluster separation 
$d_c$ in Abell cluster samples. 
The weak dependence of $r_0$ with X-ray luminosity threshold as derived from our
 analysis is partially related to the broad relation between
$L_x$ and  richness (\cite{bh94}). The relation between mass, 
richness and X-ray luminosity is uncertain and is affected by several
observational biases and systematics (contamination by projection, 
departures from hydrostatic equilibrium, etc) as well as astrophysical issues 
(galaxy formation and evolution in clusters,
pre-heating of the intra-cluster gas, shocks and supernova heating, etc).
These effects are important for
a suitable interpretation of the observations
given the different mass dependence of the cluster correlation length expected
in the variety of scenarios for structure formation. 
On the theoretical side the situation is also unclear. 
In hierarchical models of the CDM type  
the dependence of $r_0$ on $d_c$ is found either very weak (\cite{ce}), or moderate
(\cite{bc92}), discrepancies that according to \cite{ek96} may rely on the different
cluster identification algorithms. 
These considerations and the results of our analysis suggest that
cluster X-ray luminosities may be used for a reliable  
confrontation of models and observations.

\acknowledgments
 
This work was partially supported by the Consejo de Investigaciones Cient\'{\i}ficas
y T\'ecnicas de la Rep\'ublica Argentina, CONICET, the Consejo de
Investigaciones Cient\'{\i}ficas y Tecnol\'ogicas de la Provincia de C\'ordoba, CONICOR
and Fundaci\'on Antorchas, Argentina.

 {}

\end{document}